\documentstyle[twoside,epsf]{article}
\input ibvs2.sty
\input ibvst.sty
\begin{document}
	\IBVShead{}{}
	
	\IBVStitletl{NY Her: possible discovery of negative superhumps}

	\IBVSauth{Sosnovskij, A.$^{1}$; Pavlenko, E.$^{1}$; Pit, N.$^{1}$; Antoniuk, K.$^{1}$}
	
	\IBVSinst{ Crimean Astrophysical Observatory, Nauchniy, Crimea, Russian Federation, 298409. 
		$^\dagger$demartin@ukr.net, eppavlenko@gmail.com %email 
	}
	
	\SIMBADobj{NY Her}
	\IBVSkey{NY Her, dwarf novae, cataclysmic variables, photometry}

	\IBVSabs {We presents result of CDD photometry for SU UMa dwarf nova NY Her during 6 nights in June 2017 when object was in quiescence.}
	\IBVSabs {Light curves clearly show strong amplitude variations in a range of 0\fmm7-1\fmm1. Time series analysis revealed a period 0.07141(5) d,}
	\IBVSabs {that we identified as the period of possible negative superhumps of NY Her.}

	\begintext
	
	\section{Introduction}
	
		Cataclysmic variables (CVs) are composed of a white dwarf (WD) as the primary star and a Roche-lobe filling red (or brown) dwarf as the secondary star which supplies matter from the inner Lagrangian point. This matter forms an accretion disc around the primary star in the case of a non-magnetic white dwarf. The accretion disc is the main source of variability on large time intervals from minutes to hundreds of days. SU UMa-type dwarf novae are a class of CVs showing two types of outbursts: superoutbursts and normal outbursts with amplitudes of 2\fmm0-8\fmm0 (Warner, 1995).  
	
	During superoutburst these objects exhibit light variations called "positive superhumps" (Osaki, 1996). The observed period of the superhumps is a few percent longer than the orbital period of the system. 
	On the other hand, some SU UMa stars show variations shorter than the orbital period, that are called "negative superhumps" (Hellier, 2001), visible mostly in quiescence and in some occasions in the normal outbursts and superoutbursts (Harvey et al. 1995, Pavlenko et al., 2010, Oshima et al. 2014). 
	
	NY Her ($\alpha=$17:52:52.60 $\delta=$+29:22:18.8) was originally discovered by Hoffmeister (1949) as a Mira-type variable. Kato et al. (2013a) identified this object as the SU UMa-type dwarf nova with a short supercycle.  Using superoutburst data taken by the ASASSN team, Poiner's observations and results of follow-up international campaign, Kato at. al. (2017) revealed an updated positive superhump profile with a period of 0.075525 d and much smaller amplitude (0\fmm10 mag) than most of SU UMa-type dwarf novae with similar periods of superhumps (or orbital) have. They identified a possible supercycle of $\sim{63.5}$ d and that the duration of the superoutbursts was ~10 d. The supercycle length of $\sim{63.5}$ d is 	between the supercycle length of the ER UMa-type DN novae subclass (Hellier, 2001; Kato et al., 2013b) that is distinguished by the shortest (20-50d) supercycles and ordinary SU UMa stars which have supercycles longer than 100d. The 	superoutburst duration of 10 d is much shorter than the duration of superoutbursts seen in the ER UMa-type dwarf novae. Kato et al. (2017) noticed that NY Her may be classified as unique object with a short supercycle and a small superhump amplitude despite the relatively long $P_{sh}$ and could have the negative superhumps because of infrequent normal outbursts during relatively short supercycle. This motivated us to examine this prediction by photometric investigation of NY Her during quiescence in June 2017.  
	
	\IBVSfig{10cm}{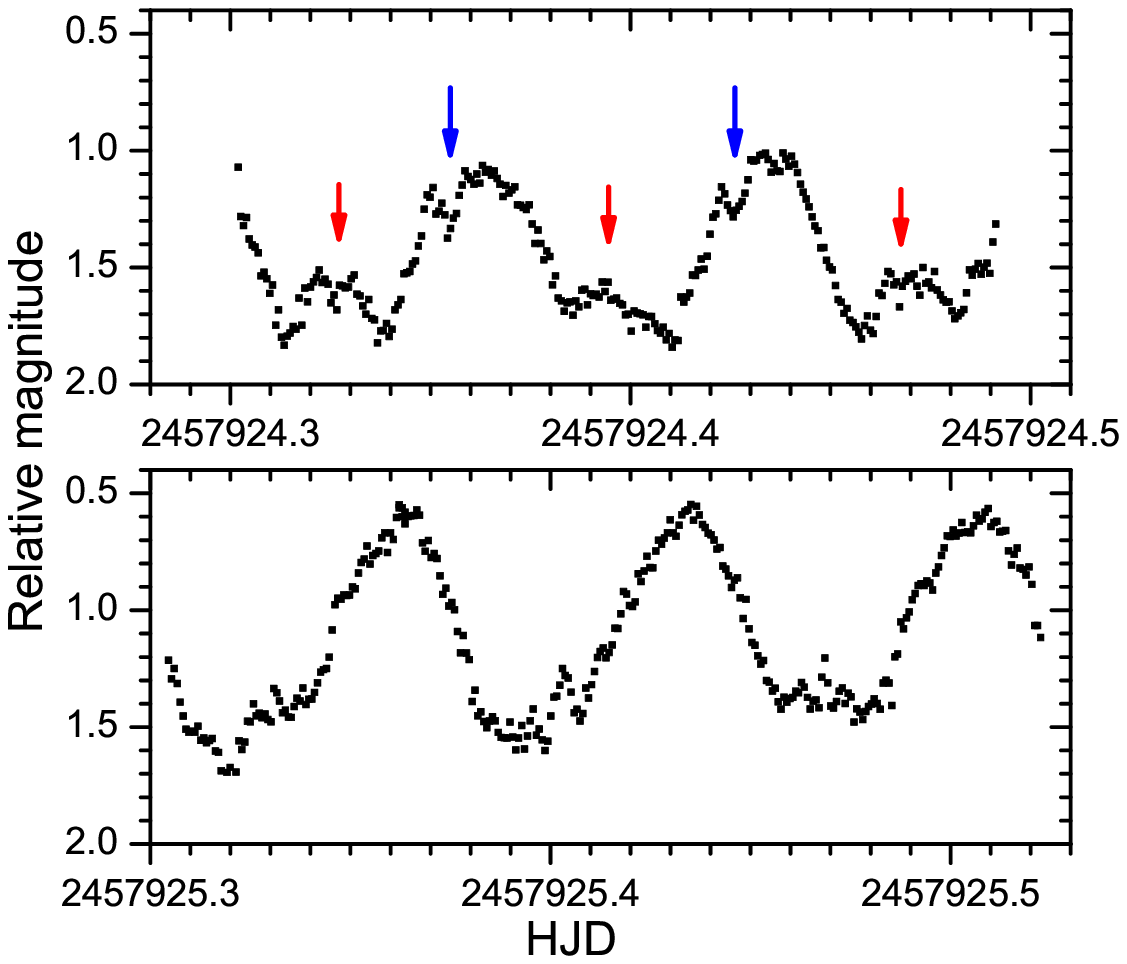}{Unfiltred photometry for NY Her for two nights: 19-20 June, 2017. The smaller humps and small dips are marked by red and blue colors correspondently. }
	\IBVSfigKey{f1.eps}{Ny Her}{photometry}
	
	\section{Observations}
	The photometric CCD observations of NY Her were carried out during 6 nights in June 2017 at the Crimean Astrophysical Observatory (CrAO) in unfiltered light, giving a system close to the $R_{c}$ band in our case, at two telescopes: 2.6--m ZTSh with APOGEE Alta E-47 and 1.25--m AZT-11 with ProLine PL230. Our priority was time series analysis with high time resolution in respect to the multicolor observations. The standard aperture photometry (de-biasing, dark subtraction and flat-fielding) was used for measuring of the variable and comparison star USNO –B1 1193-0272323 (R=17.97) (Monet et al., 2003). The accuracy of a single brightness measurement strongly depended on the telescope, exposure time, weather condition and brightness of NY Her, and reached 0\fmm01-–0\fmm03 for 60 s exposure (ZTSH) and 0\fmm08-0\fmm15 for 180 s exposure (AZT-11). 
	
	\IBVSfig{9cm}{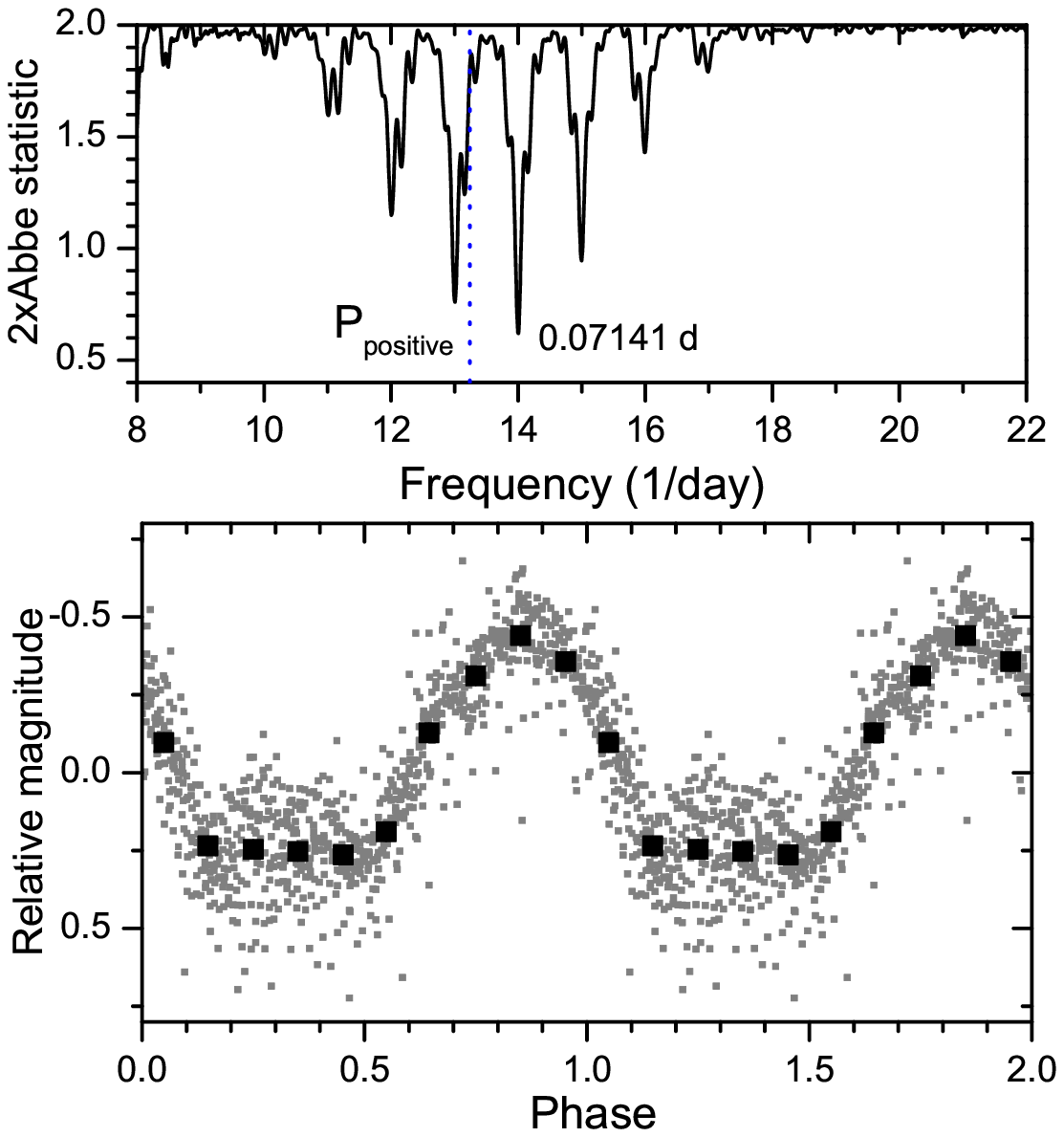}{Upper: Periodogram for combined data from 6 different nights. Position of the positive superhump period (Kato et al., 2017) is shown by blue dotted line. Lower: data folded on the 0.07141 d period. Original data are shown by gray circles. Black squares denote the mean points. }
	\IBVSfigKey{f2.eps}{Ny Her}{periodogram}

	\section{Data analysis and discussion}
	During the quiescent state the brightness of NY Her varied between 18\fmm5 and 19\fmm8. The example of two original light curves is shown in Fig.1. As seen in these light curves, the profile changes from night to night. The light curves clearly show variability with a period $\sim{1.7}$ h and strong amplitude variations in a range of 0\fmm7-1\fmm1.  At first night (Fig.1, upper frame) one could see the two humped profile with different height and small dip in bigger hump. At the second night (Fig. 1, lower frame) the light curve profiles become more smoothed, the smaller hump is no longer visible. To search for precise periodicity we have done the periodogram analysis using the Stellingwerf method (Stellingwerf, 1978) implemented in ISDA package (Pel't, 1980). The accuracy of trial periods as well as Abbe statistic, also known as Lafler-Kinman statistic (Lafler and Kinman, 1965)  was calculated using ISDA package (Pel't, 1980).  Before starting the analysis, we substracted the long term trend. The strongest peak points to the period 0.07141(5) d, surrounded by daily aliased peaks. The periodogram and phase diagram for the most significant period are shown in Fig. 2. Original data show larger scattering in minimum caused by both larger errors and intrinsic variability and smaller one in maximum. The mean light curve displays a flat minimum lasting 0.4 period and amplitude about 0\fmm7.  
	
	As empirically established relation shows, all known SU UMa stars with related Porb and Psh are located around equation line: $\epsilon$=$P_{sh}$/$P_{orb}$ -- 1 = 0.001(4) + 0.44(6)$P_{orb}$ (Kato et al., 2009). The measured period (of NY Her in quiescence) cannot be an orbital one, because in this case $\epsilon$=0.057 is situated higher this line (taking into account a scatter of observation around this line). According to this relation, the corresponding orbital period should be slightly larger, and be located in that scattering strip between 0.0722-0.0736 d, with  $\epsilon$=0.025-0.045.   
	
	We suggest that 0.07141(5) d period is the period of negative superhumps of NY Her according to Kato's prediction. However a small probability that this period could be interpreted as the orbital one also cannot be neglected  since the eclipsing SU UMa dwarf nova HT Cas has near the same large epsilon (Kato et al., 2009).  Further observations of NY Her aimed at finding the orbital period are necessary for the final identification of the brightness modulation during its quiescence in June 2017. 
	
	\emph{Acknowledgement:} We are grateful to Sklyanov A.S. from Kazan Federal University for reading ad discussion the paper and to anonymous referee for valuable comments.

	\references
	
	Harvey, D., Patterson, J.,  1995, {\it PASP},  {\bf 107}, 1055  \DOI{10.1086/133662}
	
	Hellier, C., 2001, Cataclysmic variable stars: how and why they vary, {\it Springer-Verlag London.} \DOI {ISBN 978-1-85233-211-2} 
	
	Hoffmeister, C., 1949,  {\it Erg. Astron. Nachr.}, {\bf 12}, 12 
	
	Kato, T. et al., 2009, {\it PASJ},  {\bf 61S}, 395 \DOI{10.1093/pasj/61.sp2.S395}
	
	Kato, T. et al., 2013a, {\it PASJ},  {\bf 65}, 23   \DOI{10.1093/pasj/65.1.23}
	
	Kato, T., Nogami, D., Baba, H., et al. 2013b, arXiv 1301.3202 
	
	Kato, T. et al., 2017, {\it arXiv}  {\bf 170603870} accepted to PASJ.
	
	Lafler, J. and Kinman, T.D., 1965, {\it ApJ Suppl.,} {\bf 11}, 216 \DOI{10.1086/190116}
	
	Monet, D. et al., 2003, {\it AJ}, {\bf 125},  984  \DOI{10.1086/345888}
	 
	Osaki, Y. 1996, {\it PASP}, {\bf 108}, 39 \DOI{10.1086/133689} 
	
	Ohshima, T. et al., 2014, {\it PASJ},  {\bf 66}, 670  \DOI{10.1093/pasj/psu038} 
	
	Pavlenko, E. P. et al., 2010, {\it AIPC},  {\bf 1273}, 320 \DOI{10.1063/1.3527832}
	
	Pel't, Ya., 1980  {\it Frequency Analysis of Astronomical Time Series} Tallinn:Valgus
	
	Stellingwerf, R.F., 1978, ApJ, 224, 953 \DOI{10.1086/156444}
	
	Warner, B. 1995, {\it Cataclysmic Variable Stars (Cambridge:Cambridge University Press)}

	\endreferences 
	
	%Correction: 
	
\end{document}